# Optimization of the double electron-electron resonance for C-centers in diamond

*Olga R. Rubinas[1,2,3], Vladimir V. Soshenko[2,3], Stepan V. Bolshedvorskii[2,3], Ivan S. Cojocaru[2,4], Vadim V. Vorobyov[5], Vadim N. Sorokin[2], Victor G. Vins[6], Alexander P. Yeliseev[6], Andrey N. Smolyaninov[3], Alexey V. Akimov[7,3,4*]*

[1]Moscow Institute of Physics and Technology, Dolgoprudniy, Moscow Region, Russia

[2]P. N. Lebedev Physical Institute of the Russian Academy of Sciences, Moscow, Russia

[3]LLC Sensor Spin Technologies, Moscow, Russia

[4]Rusian Quantum Center, Moscow, Russia

[5]The University of Stuttgart, Stuttgart, Germany

[6]LLC Velman, Novosibirsk, Russia

[7]Texas A&M University, 4242 TAMU, College Station, USA

E-mail: akimov@physics.tamu.edu

NV centers in diamond recommend themselves as good sensors of environmental fields as well as detectors of diamond impurities. In particular, C-centers, often also called $p_1$-centers, can be detected via double electron-electron resonance. This resonance can be used to measure the C-center concentration. Here, we measured the concentration of C-centers in several diamond plates and investigated the influence of the free precession time of the NV center on the observed contrast in the measured double electron-electron resonance spectrum. The dependence of the resonance amplitudes and widths on the concentration of C-centers as well as the length of the combined C-center driving and NV-center $\pi$-pulse is also discussed. The optimal contrast-free precession time was determined for each C-center concentration, showing a strong correlation with both the concentration of C-centers and the NV-center $T_2$ time.

## 1. Introduction

Double electron-electron resonance (DEER)[1] is often used in chemistry to determine the structure of radicals[2]. This method in its variation, based on nitrogen-vacancy (NV) centers in diamond[3], has established itself as an excellent method for detecting various spin impurities both inside diamond[4–6] and on its surface[7]. The method uses the NV center as a sensitive element for local magnetic fields[8]. The spin echo[9] signal responds to changes in the local field generated by the spin environment when a resonant radio frequency (RF) field is applied to this environment[10]. Thus, by scanning the frequency of the radio-frequency field, it is possible to obtain the DEER spectrum containing information about the possible states of the spin environment of the NV center[10].

The donor nitrogen or C-center is one of the most studied spin defects in diamond[11], since, in addition to its natural high abundance, it also limits the performance of the NV-center as a sensor. Thus, NV centers are naturally used to detect and study C centers[12]. To this end, DEER enables one to not only detect the spectrum for the defect[12] but also determine its concentration[13], which, while not inferred from the spectrum itself, can be measured using a modified pulse sequence for DEER.

In this paper, the influence of the free precession time of the NV center on the observed contrast of the measured DEER spectrum for C centers was investigated. The optimal precession time in terms of the contrast was found for each concentration of the C-center, showing a strong correlation with both the concentration of C-centers and the NV-center $T_2$ time. The concentration of C-centers (from 1 to 60 ppm) in 8 diamond plates was measured using both modified DEER spectroscopy [13] and infrared spectroscopy[14]. The dependence of the resonance amplitudes and width on the concentration of C-centers as well as the length of the combined C-center driving and NV-center $\pi$-pulse is also discussed.

## 2. Methods

All investigated diamond plates were grown by the high-pressure high-temperature method[15], irradiated with electrons[16] at a dose of $15 \cdot 10^{17} \text{cm}^{-2}$, and then annealed at a temperature of 1400°C over 2 hours. The experimental setup used for the detection of the DEER spectrum is described in detail in[13]. Experiments were conducted in an 85 G magnetic field created by permanent magnets, which allows us to split the degenerate levels of NV centers and C centers due to the Zeeman effect (Figure 1A). With this magnitude of magnetic field, the DEER spectrum of the C-center should be in the range of 100 MHz to 500 MHz. The magnetic field was set so that its direction coincided with the (111) axis of the diamond plate.

NV centers were driven at the frequency determined from the optically detectable magnetic resonance[13]. Then, using the Rabi oscillations[17] at the resonant frequency of the NV center, the duration of the π-pulse was determined, which can be controlled by the power of the microwave field applied to the ensemble of NV centers. The DEER sequence is composed of a spin echo[9] sequence on an ensemble of NV centers accompanied by a RF pulse (Figure 1B). The length of the RF pulse is chosen to match the NV $\pi$-pulse[13]. To obtain the DEER spectrum, the frequency of the RF pulse frequency is varied, while all other parameters are kept the same (Figure 1C).

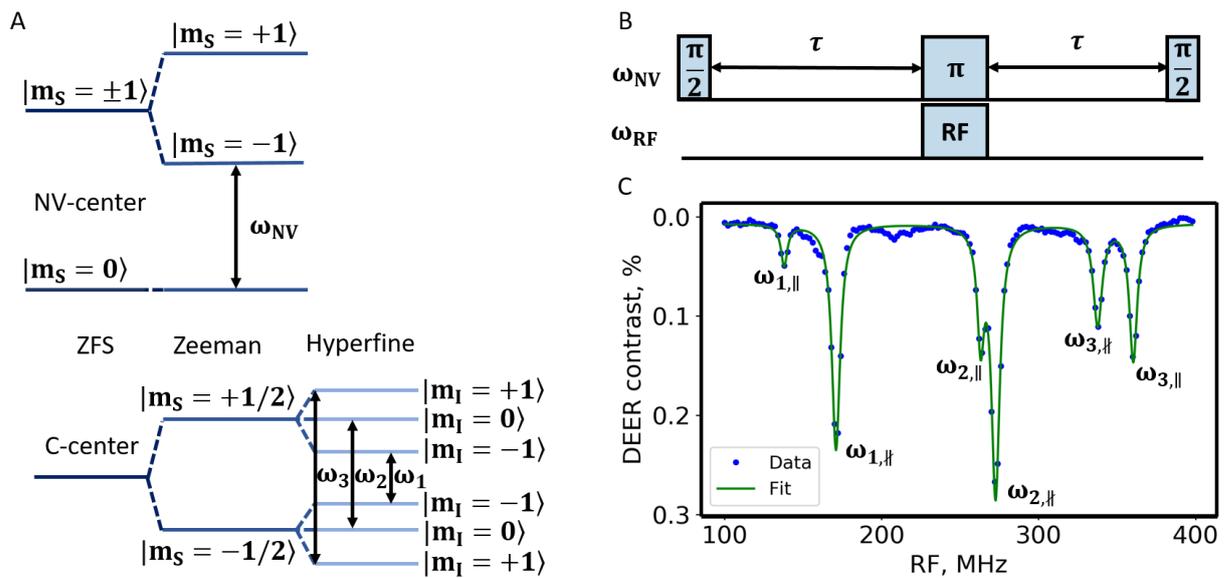

**Figure 1. A) The NV-center and C-center energy levels in a constant magnetic field. The arrows show the allowed transitions used in the experiment. B) $\omega_{NV}$ - the frequency of the NV-center transition used, $\omega_{RF}$ - the scanned RF frequency for C-center detection. C) DEER spectrum. The frequencies $\omega_{1-3}$ denote the resonances corresponding to the transitions at these frequencies in Figure 1A. The signs ∥ and ∦ denote the resonances of the C-centers that are aligned and not aligned with the external magnetic field, respectively.**

The free precession time $\tau$ was kept fixed for each specific measurement, but it is a parameter that requires optimization depending on the nitrogen concentration. Since the NV center serves as a sensor for the spectral measurements, it is expected that the properties of the NV centers can have an effect on the measured spectrum. In particular, the free precession time is limited by the coherence time $T_2$ of the NV center. The coherence time is known to correlate with the nitrogen concentration[18]. For diamond plates under study, the coherence time varied between a few microseconds for highly concentrated diamond (approximately 50 ppm of nitrogen) and 130 μs for low concentrated diamond plates (approximately 1 ppm of nitrogen). Longer $T_2$ times of up to 300 μs are possible for even lower nitrogen concentrations[19].

To measure the C-center concentration in diamonds, we used the method based on the modified DEER pulse sequence, as described in previous work[13]. To reconcile the results, the C-center concentration in these samples was also determined by infrared (IR) spectroscopy. The measurement results obtained using these methods are in good agreement (see Figure 2A). The measurement results obtained using the two methods do not coincide completely due to the inhomogeneity of the defect concentration in the diamond plates and the different investigation volumes used for the two methods. The IR spectroscopy method averages the concentration over the whole plate, whereas the DEER method obtains the concentration from the focal spot volume. To overcome concentration fluctuations, several spots were measured by DEER methods.

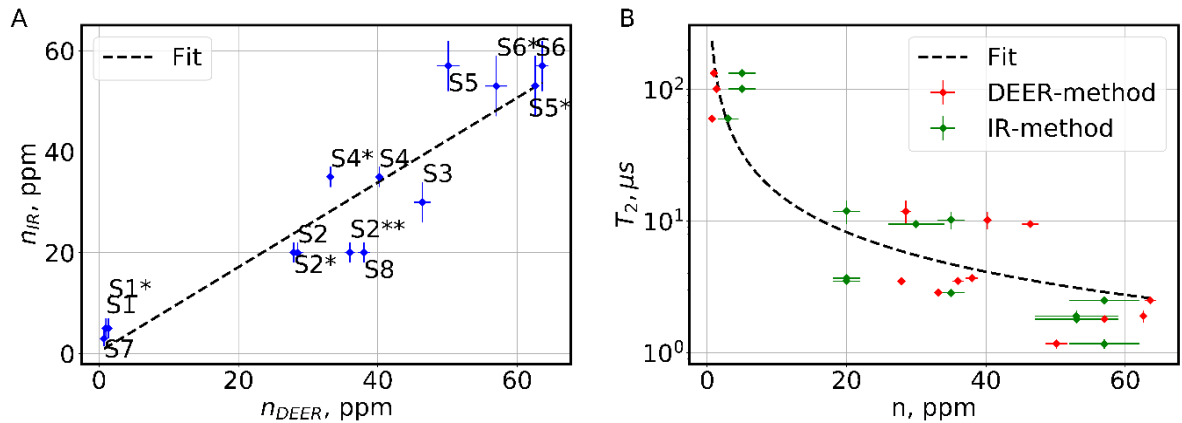

**Figure 2. A) Correlation between IR spectroscopy and DEER method measurements of C-center concentrations. $n_{IR}$ stands for the concentration measured by the IR method, and $n_{DEER}$ stands for the concentration measured by the DEER method. The dashed line demonstrates a linear fit $n_{IR}[\text{ppm}] = 0.8 n_{DEER}[\text{ppm}] + 0.3[\text{ppm}]$. B) $T_2$ coherence time of the NV-center ensemble depending on the C-center concentration. The dashed line shows the empirical estimation $T_2[\mu s] = 165[\mu s \cdot \text{ppm}]/n[\text{ppm}]$ from[18].**

The coherence time $T_2$ was measured for all samples studied using the spin-echo method[20]. Figure 3A demonstrates the measurement of $T_2$ for the plate with a nitrogen concentration of

$50\pm1.5$ ppm. Plates with low nitrogen concentrations experience modulation of the echo sequence by $^{13}$C nuclear spin[14], as shown in Figure 3B. The measured coherence times in our experiment also correlate with the measured values for the C-center concentrations, as shown in Figure 2B, and indeed follows the dependence $T_2[\mu s] = 165[\mu s\cdot ppm]/n[ppm]$ from [18]. Additionally, $T_2$ sets the upper limit on the possible free precession time $\tau$, thus affecting the parameters for the observed DEER spectra (Figure 3C,D).

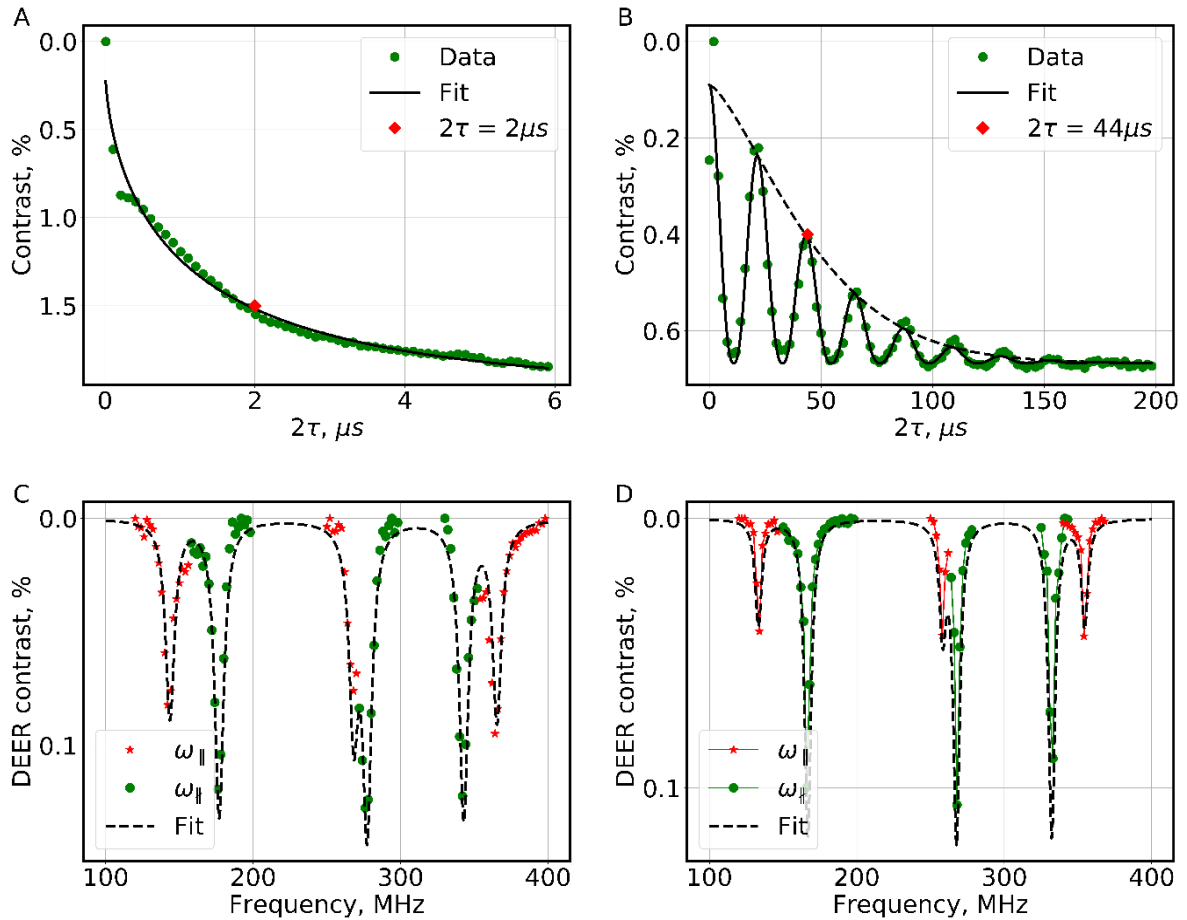

**Figure 3. NV-centered spin-echo signal for diamond with a concentration of approximately 50 ppm A) and a concentration of approximately 1 ppm B). Partial optical DEER spectra obtained for diamond with C) the diamond plate from A and free precession time $\tau = 1\ \mu s$ (red dot at A) and D) with the diamond plate from B and free precession time $\tau = 22\ \mu s$ (red dot at B). For C and D, red dots and lines correspond to the part of the C-center ensemble aligned along the magnetic field, and the green dots and lines correspond to the part that is not aligned with the magnetic field. RF power was adjusted for each resonance so that the RF pulse was a $\pi$-pulse.**

## 3. Results

The free precession time $\tau$ has several effects on the observed DEER spectra. To better understand the spectrum parameters, the measured spectra were fitted by the Lorentz fitting function:

$$L(\omega) = 1 - \sum_{i=1}^{6} C_i \frac{\Delta\omega_i^2}{(\omega - \omega_i)^2 + \Delta\omega_i^2}, \tag{1}$$

where $C_i$ – contrast, $\Delta\omega_i$ – half-width at half maximum (HWHM) of the resonance peak, and $\omega_i$ – resonance frequency are fit parameters. With such a fit, for an optical DEER spectrum measured at ~50 ppm (Figure 3C) C-centers ensemble, the HWHM on average is 3.7 MHz and that for a 1 ppm ensemble (Figure 3D) on average is approximated as 2.3 MHz, which, thus, shows a change in the HWHM. Here, to take into account the efficiency of the antenna and differences in oscillation strengths, RF power was adjusted so that the RF pulse area is equal to $\pi$ for different resonances. Therefore, the spectrum was taken piecewise[13]. The relative change in linewidth is much smaller than both the change in concentration and the change in free precision time. Thus, in the range of the investigated concentrations, the linewidth is not defined by concentration and is not Fourier limited by the free precession time. A similar conclusion was previously reported [21], where inhomogeneous broadening of C-centers is believed to be the main contribution to the resonance widths.

Next, the effect of free precession time $\tau$ on the DEER spectrum was investigated. This time was varied in the range from 200 ns to a value not exceeding twice the coherence time in each sample studied. It was observed that the contrast $C_i$ in (1) of the DEER spectrum varies in magnitude at different values of $\tau$, as shown in the attached Figure **4**A. It is easy to see that all six presented resonances (marked with different colors) have the maximum contrast at the same optimal free precession time. To find the value for this optimal free precession time, the dependence of the DEER contrast on $\tau$ was approximated by the following function:

$$F(x) = Axe^{-\left(\frac{x}{\sigma}\right)^{\beta}}, \tag{2}$$

where $A$, $\beta$, $\sigma$ are fitting parameters. This function was differentiated to find the point of maximum – optimal $\tau$. As shown in Figure **4**A, the optimal $\tau$ depends on the concentration of C-centers in the sample. A comparison of the measured $T_2$ time and optimal $\tau$ is provided in Figure 4B. The optimal free precession time is consistent with the linear dependence on the NV center coherence time $T_2$.

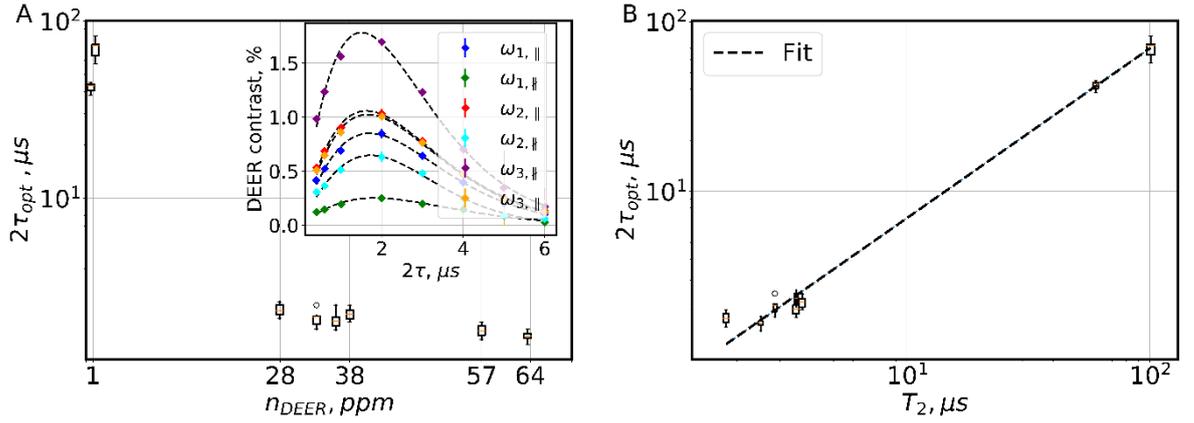

**Figure 4. A)** Concentration dependence of optimal free precession time $\tau_{opt}$ obtained from DEER. Boxes are used to show the variation in the optimal free precession time between resonances of one sample. The inset shows the dependence of the DEER contrast on the free precession time for each resonance of a single sample. **B)** Correlation between optimal free precession time $\tau_{opt}$ and NV-center coherence time $T_2$. Boxes denote experimental data. The dashed line denotes the fit to the linear function $\tau_{opt} = 0.35 T_2$.

In the next part of the experiment, RF and NV $\pi$-pulse durations in the DEER pulse scheme (Figure 1B) were varied. For each duration, the microwave power for the NV centers and the RF power for the C centers were tuned by observing Rabi oscillations to maintain pulse widths equal to $\pi$. As shown in Figure 5A, the contrast in the DEER spectrum depends on the $\pi$-pulse duration. The best contrast is achieved at the minimum $\pi$-pulse duration in the scheme. However, the linewidth of the spectrum also increases with decreasing $\pi$-pulse (Figure 5B). The dependence of this linewidth behavior on the $\pi$-pulse duration correlates with the results obtained in[21]. The width also slightly changes with nitrogen concentration, as shown in Figure 5C; this change is also rather small.

Thus, the contrast in the DEER spectrum depends on two main parameters in the experiment: the free precession time and $\pi$-pulse duration. However, the DEER linewidth weakly depends on the $\pi$-pulse duration and does not change with changes in the free precession time (Figure 5B, inset).

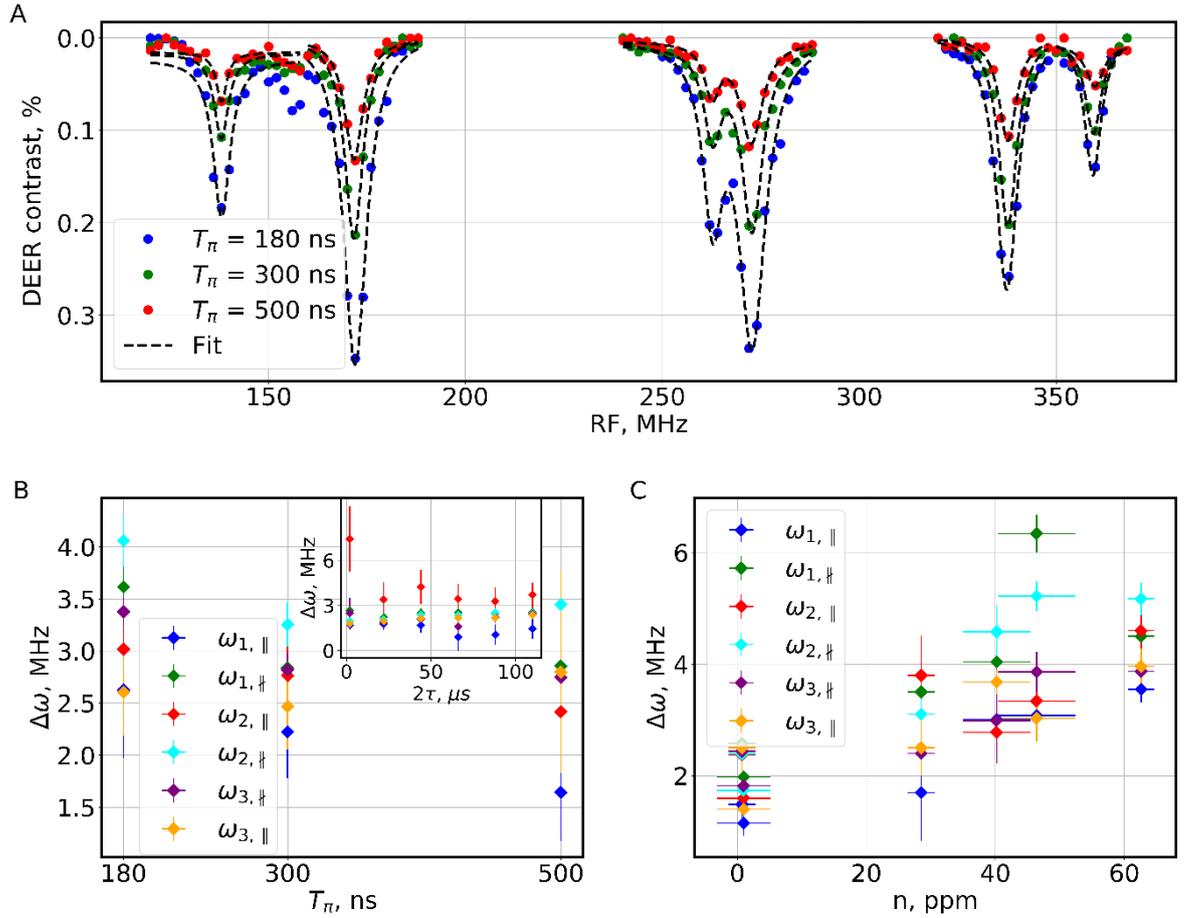

**Figure 5.** A) Piecewise DEER spectra for an ensemble of C-centers. The spectra obtained at different π-pulse durations $T_\pi$ are shown by different colors. The dashed line shows the fit to the Lorentz function. B) Dependence of the DEER spectrum linewidth on the π-pulse duration. The six resonances of the spectrum are shown by different colors. The error is the error in the approximation of the experimental data using the Lorentz function. C) Dependence of the DEER spectrum linewidth on the C-center concentration in the sample. The concentration was determined by using the modified DEER method.

## 4. Discussion

The fact that the linewidth of the DEER resonance barely deepens for all experimental parameters can be explained by the assumption that the linewidth for the C-center is limited by the inhomogeneous broadening of the C-center ensemble itself. This assumption agrees with previous study[21]. The fact that the width does not change much with the concentration of C-centers suggests that the dynamics of the C-centers are weakly affected by their concentration in the studied range of concentrations. At the same time, the strong correlation of the DEER contrast with the optimal free-precession time and the correlation of the latter with the NV-center $T_2$ time suggest that the contrast in the DEER spectrum is limited only by NV-center performance, and not by the dynamics of the C-center other than the effect of these dynamics on the NV-centers themselves.

## 5. Conclusion

We considered in detail the method of optical double electron-electron resonance in diamond using the example of detecting ensembles of C-centers through the fluorescence emitted from an ensemble of NV-centers. The influence of the duration of free precession in the optical DEER scheme on spectral contrast was investigated using samples of both highly concentrated and low-concentrated diamond plates. Using these dependencies, the optimal values for the duration of the free precession time in the optical DEER experiment were determined. We also demonstrated how the concentration of C-centers in a diamond plate affects the linewidth of the optical DEER spectrum.

The concentrations of C-centers in a number of diamond plates were measured using a modified DEER sequence and compared with concentrations measured by infrared spectroscopy. The dependence of the DEER spectra on the parameters of the DEER sequence was investigated. It was shown that the most sensitive parameter for the DEER spectrum is its contrast, and that this strongly depends on both the free-precession time and length of the $\pi$-pulse of the sequence. The free precession time has an optimal value in terms of the DEER spectrum contrast and is linearly dependent on the NV-center $T_2$ time. The width of the resonance has a weak dependence on the parameters of the sequence. The main influence of the resonance width arises from the length of the $\pi$-pulse of the sequence, but, overall, the width has other nonsequence-related sources, presumably the dynamics of the C-center themselves.

## 6. Acknowledgments

The reported study was funded by the Russian Science Foundation (grant # 21-42-04407) in part for the free spin precession time optimization and by the Russian Foundation for Basic Research (according to the research project No. 20-32-90025, Grant Aspirants) in part concerning the influence of the C-center concentration in diamond on the coherent properties and linewidth of the DEER spectra.